\documentclass{article}
\usepackage[preprint]{spconf}
\usepackage{amsmath,graphicx,url,times}
\usepackage{amssymb,acronym,xcolor,balance,booktabs}
\usepackage{color}
\usepackage{multirow}
\usepackage{lipsum}

\toappear{To appear in {\it Proc.\ WASPAA2021}}
\copyrightnotice{\copyright\ IEEE 2000}

\acrodef{STFT}{short-time Fourier transform}
\acrodef{PSD}{power spectral density}
\acrodef{RTF}{relative transfer function}
\acrodef{SNR}{signal-to-noise ratio}
\acrodef{SRR}{signal-to-reverberation ratio}
\acrodef{PDF}{probability density function}
\acrodef{DOA}{direction-of-arrival}
\acrodef{VAD}{voice activity detector}
\acrodef{MVDR}{minimum variance distortionless response}
\acrodef{LCMV}{linearly constrained minimum variance}
\acrodef{MPDR}{minimum power distortionless response}
\acrodef{LCMP}{linearly constrained minimum power}
\acrodef{AIR}{acoustic impulse response}
\acrodef{PESQ}{perceptual evaluation of speech quality}
\acrodef{CD}{cepstral distance}
\acrodef{fwSNR}{frequency-weighted segmental SNR}
\acrodef{WER}{word error rate}
\acrodef{SPP}{speech presence probability}
\acrodef{DNN}{deep neural network}
\acrodef{RNN}{recurent neural network}
\acrodef{LSTM}{long-term short-term}
\acrodef{GCC}{generalized cross-correlation}
\acrodef{RMSE}{root-mean-square error}
\acrodef{IDOA}{instantaneous direction-of-arrival}
\acrodef{SRP-PHAT}{steered response power with phase transform}
\acrodef{UCA}{uniform circular array}
\acrodef{BS-RM}{beamspace root-MUSIC}
\acrodef{CPSD}{cross-power spectral density}
\acrodef{MCLP}{multichannel linear prediction}
\acrodef{DRTF}{direct relative transfer function}
\acrodef{RETF}{relative early transfer function}
\acrodef{APA}{affine projection algorithm}
\acrodef{GSC}{generalized sidelobe canceller}
\acrodef{MAC}{multiply-accumulate}
\acrodef{CRUSE}{convolutional recurrent network for speech enhancement}
\acrodef{ASR}{automatic speech recognition}
\acrodef{MIMO}{multiple-input multi-output}
\acrodef{MISO}{multi-input single-output}
\acrodef{R-WPD}{Recursive Weighted Power minimization Distortionless response}
\acrodef{RLS-WPD}{recursive least-squares WPD}

\newcommand{\E}[1]{\operatorname{E}\left\{#1\right\}}

\def\diag{\operatorname{diag}}

\definecolor{matlab1}{rgb}{0, 0.4470, 0.7410}
\definecolor{matlab2}{rgb}{0.8500, 0.3250, 0.0980} 
\definecolor{matlab3}{rgb}{0.9290, 0.6940, 0.1250} 
\definecolor{matlab4}{rgb}{0.4940, 0.1840, 0.5560} 
\definecolor{matlab5}{rgb}{0.4660, 0.6740, 0.1880} 

\title{Low complexity online convolutional beamforming}

\name{Sebastian Braun, Ivan Tashev} 
\address{Microsoft Research, Redmond, WA, USA\\
\{sebastian.braun, ivantash\}@microsoft.com}

\begin{document}
\ninept
\maketitle

\begin{sloppy}

\begin{abstract}
Convolutional beamformers integrate the multichannel linear prediction model into beamformers, which provide good performance and optimality for joint dereverberation and noise reduction tasks. While longer filters are required to model long reverberation times, the computational burden of current online solutions grows fast with the filter length and number of microphones. In this work, we propose a low complexity convolutional beamformer using a Kalman filter derived affine projection algorithm to solve the adaptive filtering problem. The proposed solution is several orders of magnitude less complex than comparable existing solutions while slightly outperforming them on the REVERB challenge dataset.
\end{abstract}


\section{Introduction}
\label{sec:intro}
Increasing use of voice interfacing on mobile and wearable devices in diverse acoustic scenarios to communicate with machines as well as human-to-human telephony continues to pose more challenging demands on speech enhancement algorithms. 
Especially increasing distances between microphones and speech source degrade the \ac{SNR} and \ac{SRR}, which affects listener fatigue and intelligibility for humans and performance of \ac{ASR} systems \cite{Kinoshita2016}.

Multi-microphone processing enables use of spatial information in addition to spectro-temporal information, typically enabling improved enhanced speech quality and intelligibility. Multichannel speech enhancement approaches are fixed coherence beamforming \cite{Benesty2008a}, adaptive beamforming using parametric sound field models \cite{Thiergart:2014ae}, direction-only constraints \cite{Cherkassky2016}, eigenvalue decomposition \cite{Warsitz2007}, or mask-based updates \cite{Heymann2016}, additional post-filtering \cite{Cauchi2015}, and MIMO processing \cite{Nakatani2010,Braun2018a}, and combinations thereof. Frequency-domain multi-frame \ac{MIMO} filtering based on the \ac{MCLP} model \cite{Nakatani2010,Braun2018a,Jukic2017,Dietzen2017} has been proven very effective to reduce reverberation. The \ac{MCLP} model has recently been integrated into beamformers \cite{Nakatani2019,Dietzen2020,Hashemgeloogerdi2020}, which results in less complex \ac{MISO} systems while ensuring optimality. These systems are also referred to as \emph{convolutional beamformers}, and are subject to this study.
Online processing systems are more practical as they enable use of the same system for both real-time communication and low-delay \ac{ASR}. While there exist several online processing convolutional beamformers \cite{Nakatani2019a,Dietzen2020,Hashemgeloogerdi2020}, computational complexity can be still high when targeting implementation on resource-constrained devices.

In this work, we propose a low-complexity \ac{APA} solution to the convolutional beamformer, which is derived from our previous Kalman filter solution \cite{Hashemgeloogerdi2020}. The proposed system is essentially an optimal integration of a \ac{MPDR} beamformer with a reverberation canceller. 
In contrast to \ac{GSC}-based solutions \cite{Dietzen2020,Nakatani2019a,Dietzen2019}, the proposed constrained filter suffers less from signal cancellation, as it does not require a blocking matrix, whose orthogonality assumption is often violated in practice.

We show that the proposed \ac{APA} beamformer solution is by several orders of magnitudes less complex than the related \ac{R-WPD} and \ac{RLS-WPD} beamformers \cite{Nakatani2019a}. 
For the proposed convolutional \ac{APA} beamformer, we propose a zero-complexity integrated speech \ac{PSD} estimation and an optional \ac{DNN}-based \ac{PSD} enhancement. We propose an additional simplification of the convolutional \ac{APA} beamformer assuming a fixed noise field coherence.
The proposed low-complexity solution achieves comparable \ac{ASR} results to the best online system in \cite{Nakatani2019a}, which additionally relies on complex \ac{MIMO} pre-processing and a \ac{DNN} for steering vector estimation, while our steering vector is based on a low-complexity localization system \cite{Braun2019a}.
This paper distinguishes from our previous work \cite{Hashemgeloogerdi2020} by the reduced complexity adaptive filter, the fixed coherence convolutional beamformer, improved \ac{PSD} estimators, complexity and \ac{ASR} analysis.

\section{Signal Model}
\label{sec: sigmodel}
We assume $M$ microphones capturing the sound in a reverberant and noisy environment. The $m^{th}$ microphone signal in the STFT domain is denoted by $Y_m (k, n)$, where $k$ and $n$ are the frequency and time indices, respectively. The vector of microphone signals $\mathbf{y}(k, n) = [Y_1(k,n), \cdots, Y_M(k,n)]^T$ is modeled by
\begin{equation}
	\mathbf{y}(k,n) = \mathbf{a}(k,n) X(k,n) + \mathbf{r}(k,n) + \mathbf{v}(k,n),
\end{equation}
where $X(k,n)$ is the desired speech signal at the reference microphone, $\mathbf{a}(k,n)$ is the acoustic \ac{RTF} vector, and $\mathbf{r}(k,n)$ and $\mathbf{v}(k,n)$ denote reverberation and additive noise, respectively. 

The late reverberation can be modeled using the \ac{MCLP} model \cite{Nakatani2010} as a by $D$ delayed prediction from the past $L$ frames in each frequency band by
\begin{equation}
	\label{eq:MCLP_model}
	\mathbf{r}(k,n)= \sum_{l=D}^{L}\mathbf{C}_l(k,n)\mathbf{y}(k,n-l),
\end{equation}
where the matrices $\mathbf{C}_l(k,n), \, l\in\{D,\hdots,L\}$ denote the \ac{MCLP} coefficients, and $L> D\geq 1$. Note that strictly speaking, the \ac{MCLP} model \eqref{eq:MCLP_model} is only valid when the noise contribution $\mathbf{v}(k,n)$ vanishes, i.e. at higher \ac{SNR}.
%
The frequency index $k$ is omitted in the rest of the paper for better readability.

\section{Joint minimum power beamforming and multichannel linear prediction}
\label{sec:mvdr-apa}
In this section, we propose a method for joint adaptive beamforming and reverberation cancellation at the beamformer output. 
%
The desired signal $X(n)$ is estimated by obtaining the beamformer output $X_\text{b}(n)$ for the current frame, and subtracting the reverberation at the beamformer output $X_\text{r}(n)$ predicted from past $L$ frames by
\begin{equation}
	\label{eq:filtered_signal}
	\widehat{X}(n) = \underbrace{\mathbf{w}_\text{b}^T(n)\mathbf{y}(n)}_{X_\text{b}(n)} - \underbrace{\sum_{l=D}^{L}\mathbf{c}_l^T(n)\ \mathbf{y}(n-l)}_{X_\text{r}(n)}
\end{equation}
where $\mathbf{w}_\text{b}$ are the beamformer coefficients, and $\mathbf{c}_{\text{r},l}(n), \, l\in\{D,\hdots,L\}$ are the prediction filters of the reverberation at the beamformer output, obtained from the \ac{MCLP} model \eqref{eq:MCLP_model}. 
To obtain a compact vector notation, \eqref{eq:filtered_signal} is re-written as
\begin{equation}
	\label{eq:x_stacked}
	\widehat{X}(n) = \mathbf{w}^T(n) \, \mathbf{\tilde{y}}(n)
\end{equation}
where 
$\mathbf{\tilde{y}}(n) = [\mathbf{y}^T(n), \mathbf{y}^T(n\!-\!D), \hdots, \mathbf{y}^T(n\!-\!L)]^T$ are stacked microphone signals, and $\mathbf{w}(n) = [\mathbf{w}_\text{b}^T(n), -\mathbf{c}_D^T(n), \hdots, -\mathbf{c}_L^T(n)]^T$ are stacked beamformer and reverberation prediction coefficient vectors of length $Q = M(L\!-\!D\!+\!2)$, respectively.

\subsection{Constrained Kalman filter beamformer}
The joint convolutional beamformer weights $\mathbf{w}(n)$ are obtained by minimizing the power of the output signal $\widehat{X}(n)$ under the directional beam-steering constraint, generally known as \ac{MPDR} beamformer, i.\,e.,
\begin{equation}
	\label{eq:minization}
	\underset{\mathbf{w}}{\arg\min} \E{| \mathbf{w}^T \, \mathbf{\tilde{y}}|^2} \;\text{s. t.}\; 
	\mathbf{w}^{T}\mathbf{\tilde{a}} + \epsilon_a = 1,
\end{equation}
where $\mathbf{\tilde{a}} = [\mathbf{a}^T,\; \mathbf{0}_{1\times M(L-D+1)}]^T$ is the zero-padded \ac{RTF} vector. The directional constraint in \eqref{eq:minization} is relaxed by introducing the small additive error $\epsilon_a(n)$, which can model inaccuracies between the estimated and true steering vector.

We reformulate the observation into a two-equation system \cite{Chen1993,Cherkassky2016}, where the first row is obtained by re-arranging \eqref{eq:x_stacked}, and the second row is the directional constraint from \eqref{eq:minization}, i.\,e.,
\begin{equation}
	\label{eq:measurement_eq}
	\underbrace{\begin{bmatrix}  0 \\ 1 \end{bmatrix}}_\mathbf{d} = 
	\underbrace{\begin{bmatrix}\mathbf{\tilde{y}}^T(n) \\ \mathbf{\tilde{a}}^T(n)  \end{bmatrix}}_{\mathbf{F}(n)} \mathbf{w}(n) + \underbrace{\begin{bmatrix} -\widehat{X}(n) \\ \epsilon_a(n) \end{bmatrix}}_{\boldsymbol{\epsilon}(n)}.
\end{equation}
By assuming $\widehat{X}(n)$ and the directional error $\epsilon_a(n)$ to be independent random variables, the observation error correlation matrix $\boldsymbol{\Phi}_{\epsilon}(n) = \E{\boldsymbol{\epsilon}(n) \boldsymbol{\epsilon}^H(n)}$ is a diagonal matrix given by
\begin{equation}
	\label{eq:Phi_epsilon}
	\boldsymbol{\Phi}_{\epsilon}(n) = \diag\left\{ \phi_X(n), \phi_a(n) \right\},
\end{equation}
where $\phi_X(n)$ and $\phi_a(n)$ are the \acp{PSD} of $\widehat{X}(n)$ and $\epsilon_a(n)$, and $\diag\{\}$ constructs a matrix with its arguments on the main diagonal and zeros elsewhere.
The unknown evolution of the time-varying filter $\mathbf{w}(n)$ can be modeled as first-order Markov process
\begin{equation}
	\label{eq:markov_model}
	\mathbf{w} (n) = \mathbf{w}(n-1) + \mathbf{q}(n), 
\end{equation}
where the independent random variable $\mathbf{q}(n)$ models the filter uncertainty over time. 
The observation equation \eqref{eq:measurement_eq} and state model \eqref{eq:markov_model} lead to the Kalman filter solution described in \cite{Hashemgeloogerdi2020}, which requires estimation of the filter error covariance 
\begin{equation}
	\mathbf{\Phi_w}(n) = \E{\left[\mathbf{w}(n)-\mathbf{\widehat{w}}(n)\right] \left[\mathbf{w}(n)-\mathbf{\widehat{w}}(n)\right]^H},
\end{equation}
where $\mathbf{\widehat{w}}(n)$ is the estimated filter.

\subsection{Low complexity adaptive filter solution}
\label{sec:lcmv_apa}
Since the full Kalman filter \cite{Hashemgeloogerdi2020} requires computing expensive updates of $\mathbf{\Phi_w}(n)$, we assume $\mathbf{\Phi_w}$ to be a fixed diagonal matrix, which simplifies the Kalman filter to kind of a regularized \ac{APA}. The recursive filter update is then obtained by
\begin{align} 
	\mathbf{K}(n) &= \mathbf{\Phi_w}\mathbf{F}^H(n) \left[ \mathbf{F}(n)\mathbf{\Phi_w}\mathbf{F}^H(n) + \mathbf{\Phi}_\epsilon(n) \right]^{-1} \label{eq:kalman_gain} \\ 
	\mathbf{\widehat{w}}(n) &= \mathbf{\widehat{w}}(n\!-\!1) + \mathbf{K}(n) \left[\mathbf{d} - \mathbf{F}(n)\mathbf{\widehat{w}}(n\!-\!1) \right] \label{eq:filter_update}
\end{align}
where $\mathbf{K}(n)$ is the Kalman gain, and $\mathbf{d}(n), \mathbf{K}(n)$ are defined in \eqref{eq:measurement_eq}. Note that most matrix multiplications in \eqref{eq:kalman_gain} can be implemented by simple element-wise operations due to the diagonality of $\mathbf{\Phi_w}$.

After the beamformer update, we obtain the final output signal by adding more control to the reverb canceller in \eqref{eq:filtered_signal}, avoiding magnitude over-subtraction of the reverberation to avoid echo-artifacts, and limiting the amount of reverb cancellation \cite{Braun2018a}
\begin{equation}
	\widehat{X}(n) = X_\text{b}(n) - \alpha_\text{r} \min\{|\widehat X_\text{r}(n)|, |X_\text{b}(n)| \} \frac{\widehat X_\text{r}(n)}{|\widehat X_\text{r}(n)|},
\end{equation}
where $0 \leq \alpha_\text{r} \leq 1$ controls the amount of reverb reduction.

We propose to model the update of the beamformer $\mathbf{w}_\text{b}(n)$ and reverberation prediction coefficients $\mathbf{c}_\ell(n)$ separately with the time-invariant variances $\phi_\text{b}$ and $\phi_\text{r}$, respectively. The filter error covariance matrix is then given by
\begin{equation}
	\mathbf{\Phi_w} = \diag\big\{ \underbrace{\phi_\text{b}, \dots, \phi_\text{b}}_{M}, \; \underbrace{\phi_\text{r}, \hdots, \phi_\text{r}}_{M(L-D+1)} \big\}.
\end{equation}

While the \ac{APA} simplification has been proposed for the standard \ac{MPDR} beamformer in \cite{Cherkassky2016}, here we use the convolutional \ac{MPDR}, keep the parameters $\mathbf{\Phi_w}$ and $\phi_X(n)$ more general, and propose novel estimators for the speech \ac{PSD} $\phi_X(n)$ in the following.

\subsection{Speech PSD estimation}
\label{sec:PSDs}
In this section, a simple but effective estimation of the desired signal \ac{PSD} required in \eqref{eq:Phi_epsilon} is proposed, with an optional further enhancement using a \ac{DNN}, as shown in Fig.~\ref{fig:blockdiag}.
\begin{figure}[tb]
	\centering
	\includegraphics[width=0.7\columnwidth,clip,trim=290 170 340 160]{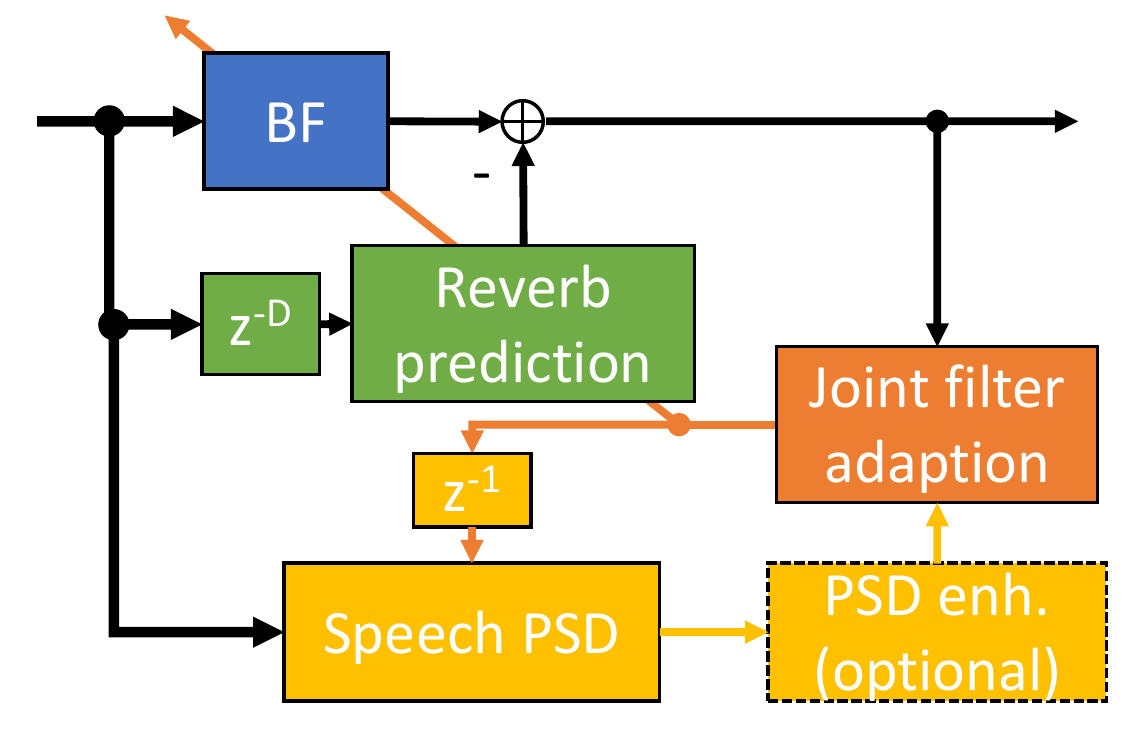}
	\caption{Proposed convolutional beamformer system. The yellow speech PSD estimation block can be replaced by a DNN. We also explore using a fixed coherence beamformer, adapting only the green reverb canceller branch.}
	\label{fig:blockdiag}
\end{figure}

The desired signal \ac{PSD} can be estimated at almost no additional cost by applying the previously estimated filter coefficients to the current frame, i.\,e.,
\begin{equation}
	\label{eq:phi_x_est}
	\widehat{\phi}_X(n)  = |\widehat{\mathbf{w}}^T(n-1) \, \tilde{\mathbf{y}}(n)|^2.
\end{equation}
Note that the filtered signal term in \eqref{eq:phi_x_est} needs to be computed in the filter update \eqref{eq:filter_update} as well, so it comes at almost no additional cost. In contrast to \cite{Hashemgeloogerdi2020}, we found any temporal smoothing or decision-directed estimation on \eqref{eq:phi_x_est} harmful to speech quality.

\textbf{DNN-based PSD enhancement:}
%
%
As we found the \ac{PSD} $\widehat{\phi}_X(n)$ to play an essential role on the beamformer performance, we propose an optional enhancement of the \ac{PSD} using a \ac{DNN} for speech enhancement as shown in Fig.~\ref{fig:blockdiag}. We use the \ac{CRUSE} proposed in \cite{Braun2021a}, which was trained to predict a spectral suppression filter to enhance single-channel recorded speech. Thus, the pre-estimated \ac{PSD} is enhanced by suppressing residual noise and reverberation by
\begin{equation}
	\label{eq:psd_dnn}
	\widehat{\phi}_X^\text{DNN}(n) = G_\text{DNN}^2(k,n) \, \widehat{\phi}_X(k,n),
\end{equation}
where $G_\text{DNN}(k,n)$ is the enhancement filter predicted by the \ac{DNN}.
The \ac{DNN} consists of 4 causal convolutional encoder and decoder layers with skip connections and a recurrent center layer. The network runs in real-time with 4.2~M MACs per frame and uses only current and past frame information. 

To mitigate speech distortion, before the \ac{PSD} estimate is inserted in \eqref{eq:Phi_epsilon}, we apply the lower bound $\max\left\{ \widehat{\phi}_X(n), \; \eta\Vert\mathbf{y}(n)\Vert^2 \right\}$ depending on the mean input signal power scaled by $\eta<1$.

\section{Fixed coherence beamformer with reverberation canceller}
\label{sec:conv-sdmvdr}
If we replace the beamformer in Fig.~\ref{fig:blockdiag} with a fixed coherence beamformer, e.g. the superdirective \ac{MVDR} \cite{Benesty2008a}, we only need to adapt the reverb canceller branch. Solving this system equivalently as in Sec.~\ref{sec:lcmv_apa} with the \ac{APA}, the observation system \eqref{eq:measurement_eq} reduces to a single equation, as we don't need the directional constraint below the first row anymore. The superdirective \ac{MVDR} beamformer obeying these directional constraints is given by
\begin{equation}
	\label{eq:sdmvdr}
	\mathbf{w}_\text{sd}(n) = \underset{\mathbf{w}}{\arg\min}\, \mathbf{w}^H \mathbf{\Gamma}_\text{d} \mathbf{w} \quad \text{s. t.}\; 
	\mathbf{w}^{H}\mathbf{a} = 1,
\end{equation}
where $\mathbf{\Gamma}_\text{d}(k)$ is the time-invariant diffuse coherence matrix that depends only and the array geometry \cite{Cron1962}. Note that the \acp{RTF} $\mathbf{a}(n)$ are in general still time-varying.
Consequently, the measurement equation for the adaptive filter becomes
\begin{equation}
	\underbrace{\mathbf{w}_\text{sd}^H(n) \mathbf{y}(n)}_{d(n)} = \mathbf{f}^T(n) \, \mathbf{w}_\text{rc}(n) + \widehat{X}(n)
\end{equation}
where $\mathbf{f}(n) = [\mathbf{y}^T(n-\!D), \hdots, \mathbf{y}^T(n\!-\!L)]^T$ and $\mathbf{w}_\text{rc}(n) = [\mathbf{c}_D^T(n), \hdots, \mathbf{c}_L^T(n)]^T$.
The beamformer output on the left-hand side now becomes a time-frequency variant constraint $d(k,n)$ for the adaptive filter, while the matrix $\mathbf{F}(n)$ in \eqref{eq:measurement_eq} reduces to the vector $\mathbf{f}^T(n)$. The \ac{APA} filter update is obtained analogous with \eqref{eq:kalman_gain}, \eqref{eq:filter_update} by replacing $\mathbf{F}$, $\mathbf{d}$ with $\mathbf{f}^T$, $d$.
The speech \ac{PSD} estimation becomes consequently
\begin{equation}
	\label{eq:phi_x_sd}
	\widehat{\phi}_X^\text{sd}(n)  = |\mathbf{w}_\text{sd}^H(n) \, \mathbf{y}(n) - \mathbf{f}^T(n) \, \mathbf{w}_\text{rc}(n-1) |^2.
\end{equation}

\section{Beamformer complexity analysis}
The complexity for all beamformers depends on the joint filter length $Q=M(L-D+2)$, where $L=0$ yields the non-convolutional standard beamformer.
\begin{figure}[tb]
	\centering
	\includegraphics[width=0.9\columnwidth,clip,trim=10 0 10 10]{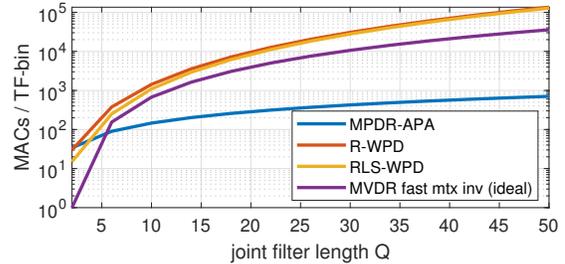}
	\caption{Complexity of APA beamformer compared to recursive WPD and RLS WPD \cite{Nakatani2019a}. Analysis involves beamformer only, excluding steering vector and neural networks.}
	\label{fig:apa_complexity}
\end{figure}
Figure~\ref{fig:apa_complexity} shows the number of \ac{MAC} operations required to compute one beamformer update per time-frequency bin for the proposed {\color{matlab1}MPDR-APA beamformer}. When the standard \ac{MVDR} \eqref{eq:sdmvdr} uses an adaptive, time-variant noise coherence estimate, such as mask-based beamformers \cite{Boeddeker2018}, it requires a matrix inversion, which we compute using the {\color{matlab4}ideally fastest matrix inversion algorithm} we could find with $O(Q^{2.37})$ \cite{Coppersmith1990}. The {\color{matlab2}\emph{R-WPD}} and {\color{matlab3}\emph{RLS-WPD}} algorithms \cite{Nakatani2019a} employ efficient recursive matrix inversion. 

We can observe that the complexity of the \ac{APA} solution grows only linearly, while the other solutions rise quadratically with $Q$, as they have to handle large full-size $Q\times Q$ matrices.
Note that the fast matrix inversion MACs are theoretical, and an equivalent speed-up might not be achieved in practical implementations, still justifying preference of the recursive WPD solutions over standard MVDR.
The proposed \ac{APA} beamformer is a computationally favorable choice for larger $Q$:
While the standard \ac{MVDR} solution requires inversion of an $Q\times Q$ matrix, the MPDR-APA requires only $2\times2$ matrix inversion in \eqref{eq:kalman_gain}.
The computational advantage pays off for setups with larger number of microphones $M>4$, and especially when using a convolutional beamformer. For an $M=8$ mic setup, the total filter length $Q$ can easily exceed 100 taps for typical convolutional filter lengths in the range $L=[6,20]$.

\section{Experimental validation}
\label{sec:results}


\subsection{Evaluation setup}
For public comparability, we show results on the REVERB challenge evaluation set \cite{Kinoshita2016}, comprising simulated and real recordings using a uniform circular 8 microphone array with radius 10~cm in reverberant rooms and moderate background noise.
The algorithms are evaluated using \ac{CD}, \ac{fwSNR}, and \ac{WER}. The \ac{WER} was obtained using the Kaldi \cite{Povey2011} REVERB challenge baseline speech recognizer using a TDNN acoustic model trained using lattice-free MMI and online i-vector extraction, and a tri-gram language model.
We also measured the runtime of the beamformers only (without steering vector estimation) in NumPy as processing time per second of audio for an 8-channel setup.

In addition to the proposed convolutional \ac{APA} beamformer (\emph{convMPDR-APA}), we also evaluate the plain \emph{MPDR-APA} beamformer without reverb canceller, which is a special case of the proposed framework for $L\!=\!0$. 
Both standard and convolutional beamformers are used without and with DNN-based \ac{PSD} enhancement described in Sec.~\ref{sec:PSDs}, where the acronym \emph{convMPDR-APA-DNN} represents the convolutional beamformer with DNN PSD enhancement. In addition, we show the superdirective (fixed coherence) beamformer with adaptive reverb canceller proposed in Sec.~\ref{sec:conv-sdmvdr}.
As baselines we have the unprocessed \emph{reference microphone}, \emph{delay\&sum} and \emph{superdirective \ac{MVDR}} beamformers, the single-channel \emph{DNN} (CRUSE) \cite{Braun2021a} applied on the reference mic, mask-based MVDR beamformer using the DNN-mask to adaptively update the noise covariance \cite{Boeddeker2018}, and the competitive \emph{RLS-WPD} \cite{Nakatani2019a} as the state-of-the-art online convolutional beamformer. 

The proposed methods are implemented with a \ac{STFT} using 50\% overlapping 32~ms square-root Hann windows and a 512-point FFT on 16~kHz sampled signals. The convolutional filter lengths are $L\!=\!\{12,8,6\}$ in three frequency bands with transition frequencies $\{800,2000\}$~Hz. The beamformer and reverb canceller filter variances are $\phi_\text{b} \!=\! -37$~dB and $\phi_\text{r} \!=\! -40$~dB. The directional uncertainty is $\phi_a \!=\! -120$~dB and the speech \ac{PSD} estimates are limited with $\eta \!=\! -25$~dB. The steering vector $\mathbf{a}(k,n)$ is estimated using a spatial probability-based far-field localization method \cite{Braun2019a} based on the simple plane wave sound propagation model.
CRUSE is trained on the data from \cite{Reddy2021} as described in \cite{Braun2021a}, only with adjusted STFT parameters.
As the test signals are very short, mostly below 10~s, all adaptive methods are initialized with a prior pass to give the adaptive algorithms a chance to converge. All parameters were tuned on the REVERB development set, and results are shown on the evaluation set.

Note that \emph{RLS-WPD} uses DNN mask-based \ac{RTF} steering vectors, while all other beamformers use the localization-based steering vectors \cite{Braun2019a}, and the results are directly obtained from \cite{Nakatani2019a}. The steering vectors for \emph{RLS-WPD} were either estimated from the mic signals, or from MIMO-WPE pre-processed signals, which is a large additional computational burden and time $T_\text{WPE}$, exceeding several times the cost of \emph{RLS-WPD} itself due to MIMO design. 


\subsection{Results}
The methods in Table~\ref{tab:results} are categorized in three groups: i) the single-channel references unprocessed microphone and \ac{DNN} only, ii) beamforming only, and iii) the proposed convolutional beamformers compared to the methods proposed in \cite{Nakatani2019a}.
\begin{table}[tb]
	\centering
	{\scriptsize
		\begin{tabular*}{\columnwidth}{l|ccc|c|c}
			\toprule
			& \multicolumn{3}{c|}{SimData} & RealData & \\
			method 	& CD & fwSNR & WER & WER & time/s \\
			\midrule\midrule
			\multicolumn{6}{c}{single-channel}\\
			\midrule
			ref mic						& 3.96	& 3.62 & 5.21	& 19.15		& 0 \\
			DNN 						& 2.94	& 8.94 & 5.74	& 16.40 	& 0.024\\
			\midrule\midrule
			\multicolumn{6}{c}{beamforming ($L\!=\!0$)}\\
			\midrule
			delay \& sum				& 3.11	& 6.37 & 4.00	& 13.11 		& 0.003\\
			superdirective MVDR			& \bf{2.98}	& \bf{6.50} & 4.00	& 13.11 & 0.004\\
			DNN-mask MVDR				& 3.14	& 6.42	 & 4.18	& 13.99 & 0.035\\
			\midrule
			MPDR-APA					& 3.99	& 3.84 & 4.18	& 13.99 	& 0.005\\
			MPDR-APA-DNN				& 3.25	& 5.98 & \bf{3.88}	& \bf{11.56} & 0.029\\
			\midrule\midrule
			\multicolumn{6}{c}{convolutional beamforming}\\
			\midrule
			RLS-WPD$^\dagger$ (no WPE)		& 3.29	& 6.08 & 4.37	& 12.80 	& 0.934\\
			RLS-WPD$^\dagger$ (+ WPE)			& 3.21	& 6.26 & 4.14	& 11.88		& 0.934+$T_\text{WPE}$ \\
			\midrule
			convMPDR-APA				& 3.79	& 4.85 & 4.79	& 12.00 & 0.009\\
			convMPDR-APA-DNN			& 2.82	& \bf{8.63} & 4.84	& \bf{10.54}  & 0.035\\
			conv-sdMVDR					& \bf{2.74}	& 7.71 & \bf{3.88}	& 10.70 & 0.007\\
			\bottomrule
		\end{tabular*}
	}
	\caption{Results on REVERB challenge evaluation dataset in terms of speech enhancement metrics, WER, and processing time.\newline $^\dagger$ are using DNN-based RTF steering vectors.}
	\label{tab:results}
\end{table}

While the \ac{DNN} is able to greatly improve speech enhancement metrics and WER in high reverberation conditions (RealData), the single-channel distortion artifacts hurt WER in the low reverberant conditions in SimData, leading to a slight overall degradation.
For non-convolutional beamformers, \emph{delay\&sum} and \emph{superdirective MVDR} outperform the adaptive \emph{MPDR-APA} and \emph{DNN-mask MVDR}, because superdirectivity is close to an optimal solution for the homogenous ambient background noise and reverberation in the REVERB dataset. In non-homogenous, time-varying noise fields, this might however change in favor of adaptive beamformers. The DNN-enhanced PSD improves the \emph{MPDR-APA} significantly, achieving the best WER for non-convolutional beamformers.

As \ac{ASR} systems are very sensitive to reverberation, the reverb cancellation of the convolutional beamformers provides a significant performance gain in terms of WER over non-convolutional beamformers. 
The DNN-based PSD enhancement provides significant gains to \emph{convMPDR-APA} especially for the speech enhancement metrics, attributed to improved noise reduction.
Finally, the convolutional superdirective beamformer \emph{conv-sdMVDR} yields comparable speech enhancement performance to \emph{MVDR-APA-DNN},with similar WER on RealData and better WER on SimData, at even lower complexity. However, we would like to stress again that the fixed noise field coherence of \emph{conv-sdMVDR} is a well fitting solution for the dataset at hand, but might not generalize as well to other noise fields including time-variant and directional noise, where the adaptive convMPDR-APA can be advantageous.

The proposed convolutional beamformers perform comparable or better in terms of speech enhancement metrics and WER compared to the state-of-the-art \emph{RLS-WPD} convolutional beamformer. Furthermore, the complexity and computation time of the proposed family of convolutional APA beamformers, even with DNN enhancement, is a fraction of the \emph{RLS-WPD} beamformer. The DNN-based steering vector estimation and MIMO-WPE preprocessing adds an additional large computational burden, where the MIMO-WPE processing time $T_\text{WPE}$ likely exceeds the \emph{RLS-WPD} time itself.

\section{Conclusion}
We proposed a scalable system integrating joint adaptive MPDR beamforming and reverberation cancellation using a Kalman filter-derived affine projection algorithm. With its complexity rising only linearly with the filter length, the proposed system is by several magnitudes less complex than previously proposed online solutions for this problem. We showed that speech PSD estimate is crucial, and be enhanced using a neural network with significant performance gains. Without any \ac{DNN}, the proposed system achieves results close to state-of-the-art systems that always rely on \ac{DNN}-based parameter estimates, while when combining our approach with a \ac{DNN}, the state-of-the-art approaches are outperformed at still lower complexity.
Further improvements from end-to-end DNN training are expected and subject to future work.

\section{Acknowledgement}
We thank Dr.\ Nakatani and his colleagues at NTT for providing their trained Kaldi baseline model to score our algorithms.

%

\balance
\bibliographystyle{IEEEtran}
\bibliography{sapref.bib}

\end{sloppy}
\end{document}